\documentclass[journal,letterpaper]{IEEEtran_nssmic}

\usepackage{graphicx}  

\begin{document}
\title{A New Scintillator Tile/Fiber Preshower Detector \\ for the CDF Central Calorimeter}
\author{Michele Gallinaro $^{\dagger *}$
\thanks{$^{\dagger}$ M. Gallinaro is with
The Rockefeller University, New York, NY 10021; phone: (212)~327-8822, fax: (212)~327-7786, email: michgall@fnal.gov.}%
\thanks{$^{*}$ representing the CDF collaboration.}
}
\maketitle
\begin{abstract}
A detector designed to measure early particle showers has been installed in front of the central CDF calorimeter at the Tevatron.
This new preshower detector is based on scintillator tiles coupled to wavelength-shifting fibers 
read out by multi-anode photomultipliers and has a total of 3,072 readout channels.
The replacement of the old gas detector was required due to an expected increase in
instantaneous luminosity of the Tevatron collider in the next few years.
Calorimeter coverage, jet energy resolution, and electron and photon identification are 
among the expected improvements.
The final detector design,
together with the R$\&$D studies that led to the choice of scintillator and fiber, 
mechanical assembly, and quality control are presented.
The detector was installed in the fall 2004 Tevatron shutdown
and started collecting colliding beam data by the end of the same year.
First measurements indicate a light yield of 12 photoelectrons/MIP, a more than two-fold increase over the design goals.
\end{abstract}


\section{Introduction}
\PARstart{T}{he} physics program at the Fermilab Tevatron collider will continue to explore the frontier of particle physics until 
the startup of the LHC experiments at CERN later in this decade.
However, detector upgrades are required beyond the ones first planned for the start of
Run~II in order to provide better detector performance for the expected increase in
instantaneous luminosity during the second phase of Run II and the associated large
data output from the future Tevatron collider program.
Calorimeter performance can be improved by replacing the old preshower gas detector 
with a new detector based on scintillator technology.
During the fall 2004 Tevatron shutdown, 
the front face of the CDF central calorimeter has been equipped with a new preshower detector based on scintillator 
tiles and read out by {\it wavelength-shifting} (WLS) fibers. Thanks to a larger signal output from the new detector
and better signal to noise ratio, both electron and photon 
identification are expected to improve significantly.

\section{Why a new detector?}
\PARstart{T}{he} replacement of the {\it Central Preshower} (CPR) and {\it Central Crack} (CCR) detectors 
with scintillator detector technology is recommended for a number of reasons:
\par\noindent 
1) The old CPR is a slow wire chamber with a relatively poor segmentation;
the instantaneous luminosity has increased since the start of Run~II by a factor of ten, and a further rise is expected 
in the next few years.
This combination would lead to high detector occupancy, jeopardizing good electron and photon identification
during the most crucial period of Run~II, when the high-energy physics frontier can be explored with large data samples
before the LHC era arrives.
\par\noindent
2) Jet energy corrections can be improved even at the current occupancy level with an expanded detector segmentation.
\par\noindent
3) The old gas detectors have been operating since the start of Run~I and wire ageing is degrading the capability of the old detector.
Furthermore, the signal pulses from a gas detector are small and signal response resolution is poor, 
when compared to those from a scintillator detector. 
Both effects are of crucial importance to electron and photon identification and energy resolution improvements.
\par\noindent
4) Electron-pion separation can also be improved using a new detector with an improved ability to detect 
{\it minimum ionizing particles} (MIPs) 
and separate them from the background of pions (see, for example, Ref.~\cite{top_prd}, p.~2996).

\vspace*{0.7cm}

\section{Extending the physics reach}
\PARstart{T}{he} CPR has already contributed to many important physics results, 
especially those involving photon and electron identification.
The upgrade of the CDF central calorimeter is expected to greatly improve the Run~II physics results and enhance the 
sensitivity to physics beyond the standard model.
In particular, it will play an important role in soft electron tagging of {\it b}-jets, photon identification, and jet resolution. 
Jet resolution can be improved by incorporating tracking information in the jet reconstruction algorithm~\cite{lami}.
Improvement of jet energy resolution can be important in enhancing the reach for new physics processes, 
such as the Higgs boson.

For the purpose of photon identification at CDF, the shower maximum detector alone cannot resolve single photons from meson decays 
(i.e. $\pi^0\rightarrow\gamma\gamma$)
above 35~GeV, as the angular separation between the two photons is too small.
Instead, the CPR uses photon conversion rates which are energy independent, and can be used at any energy range~\cite{photon_prl}.
Furthermore, it can be used to estimate the backgrounds to exotic physics signals that include photons. 
In fact, if present, new physics events are expected to appear in the high transverse momentum region
where the shower maximum has no discriminatory power.

The thin gas layers of the old CCR detectors have not been used efficiently due to several reasons, including low signal response.
However, the capability of tagging high energy electrons and photons in the crack region in events which may contain new physics 
is of paramount importance. For example, if supersymmetry manifests itself, as some suggest, in events with photons and missing transverse energy~\cite{eegg}, 
a better calorimeter coverage is highly desirable.

\section{Detector design}
\begin{figure}[h]
\centering
\includegraphics[width=3.0in]{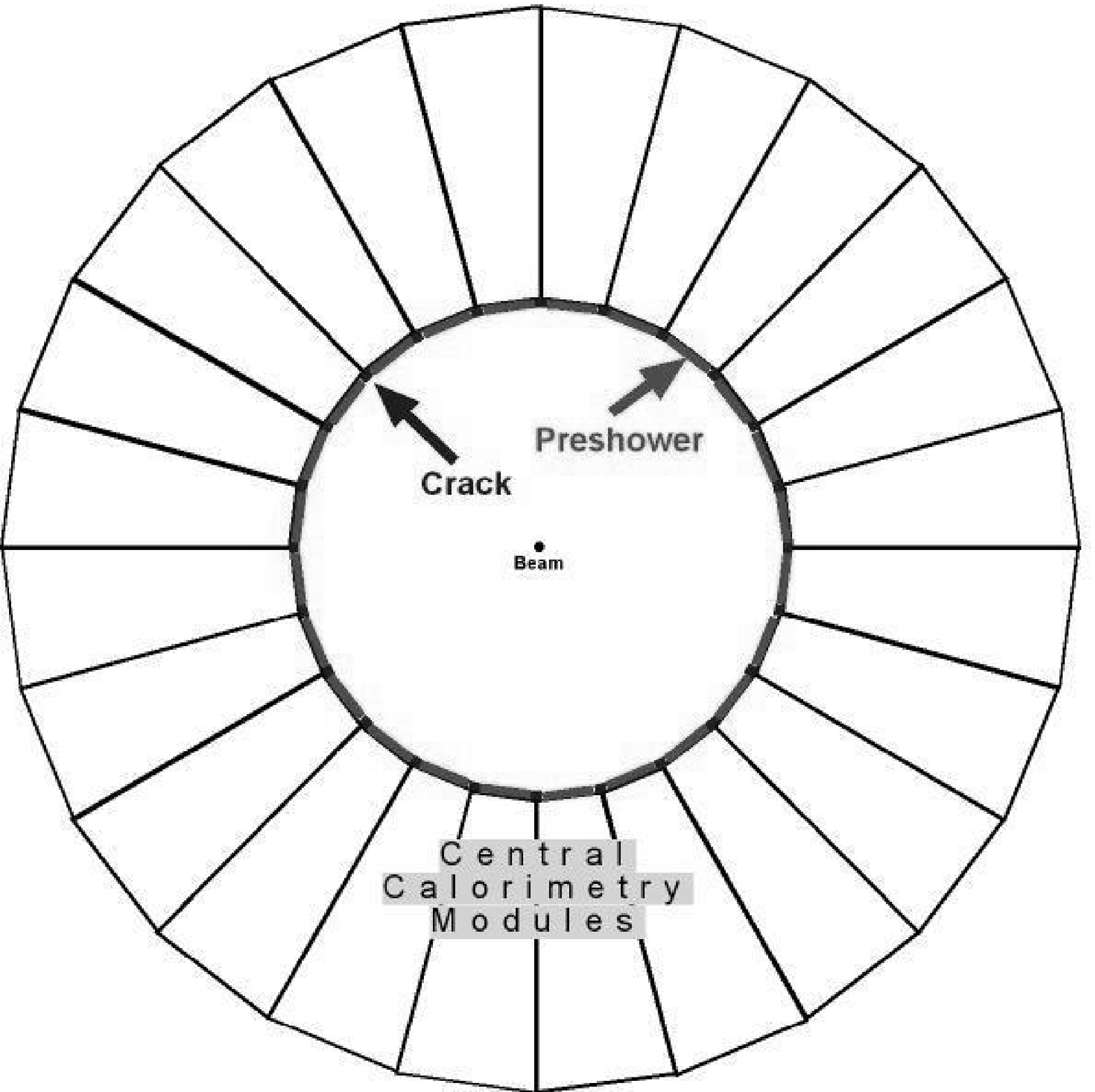}
\caption{A schematic drawing, transversal to the beam, shows
the location of the Preshower and Crack detectors, covering the front face of the calorimeter wedges.}
\label{cpr_det}
\end{figure}

\begin{figure}[h]
\centering
\includegraphics[width=3.0in]{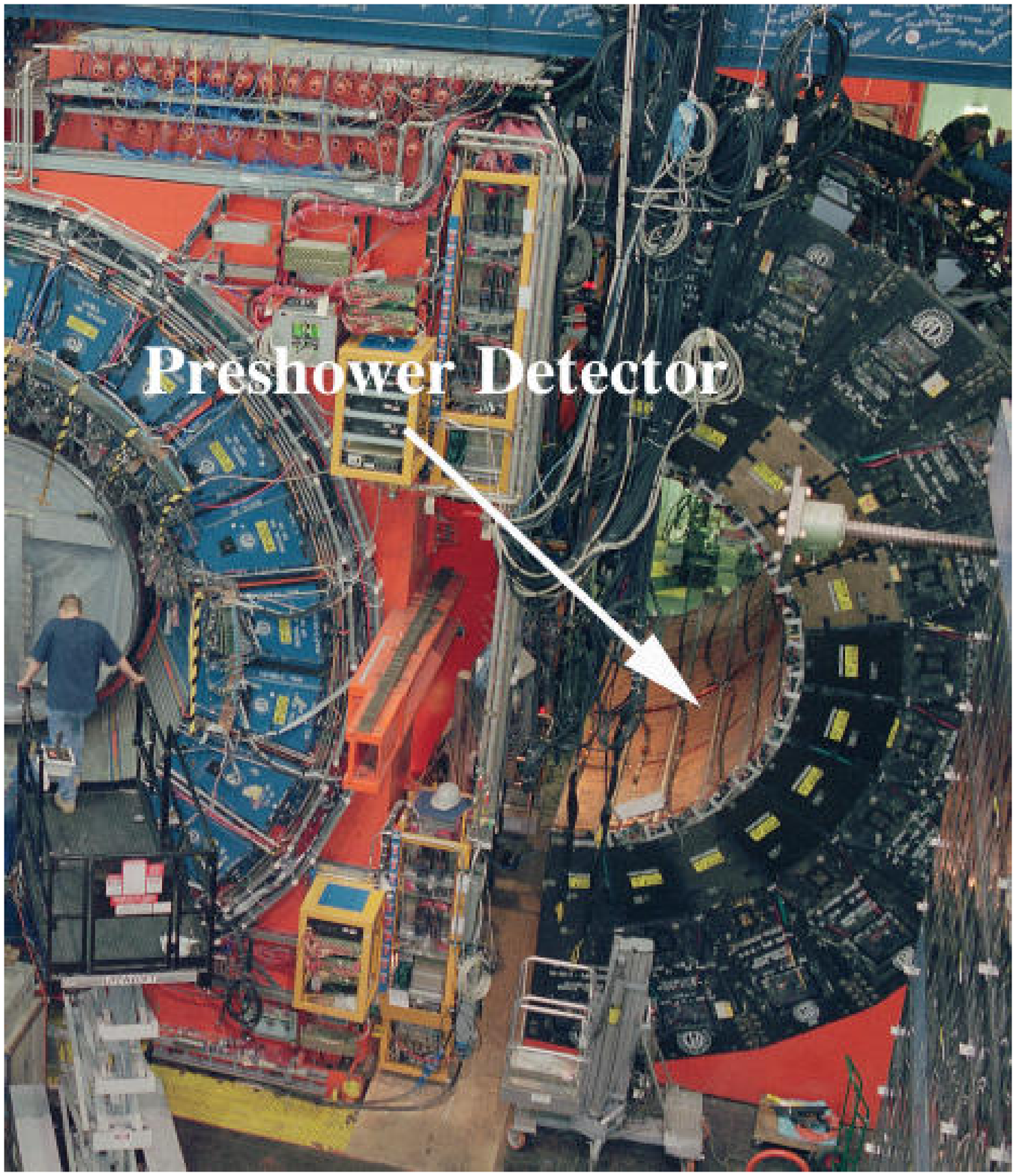}
\caption{Aerial view of a section of the CDF detector before CPR/CCR detector installation.
One calorimeter arch is open for maintenance and twelve calorimeter wedges are visible.
}
\label{cpr_det2}
\end{figure}

\PARstart{T}{he} upgrade of the CDF central calorimeter~\cite{timing} includes the replacement of 
the CPR and CCR detectors, which sample the early particle showers in front of the central calorimeter (Figs.~\ref{cpr_det} and ~\ref{cpr_det2}).
The slow gas detectors, which have been in operation since the start of Run~I, 
have just been replaced during the Tevatron shutdown in the fall of 2004, with a faster scintillator version and better segmentation.

\begin{figure}[h]
\centering
\includegraphics[width=3.0in]{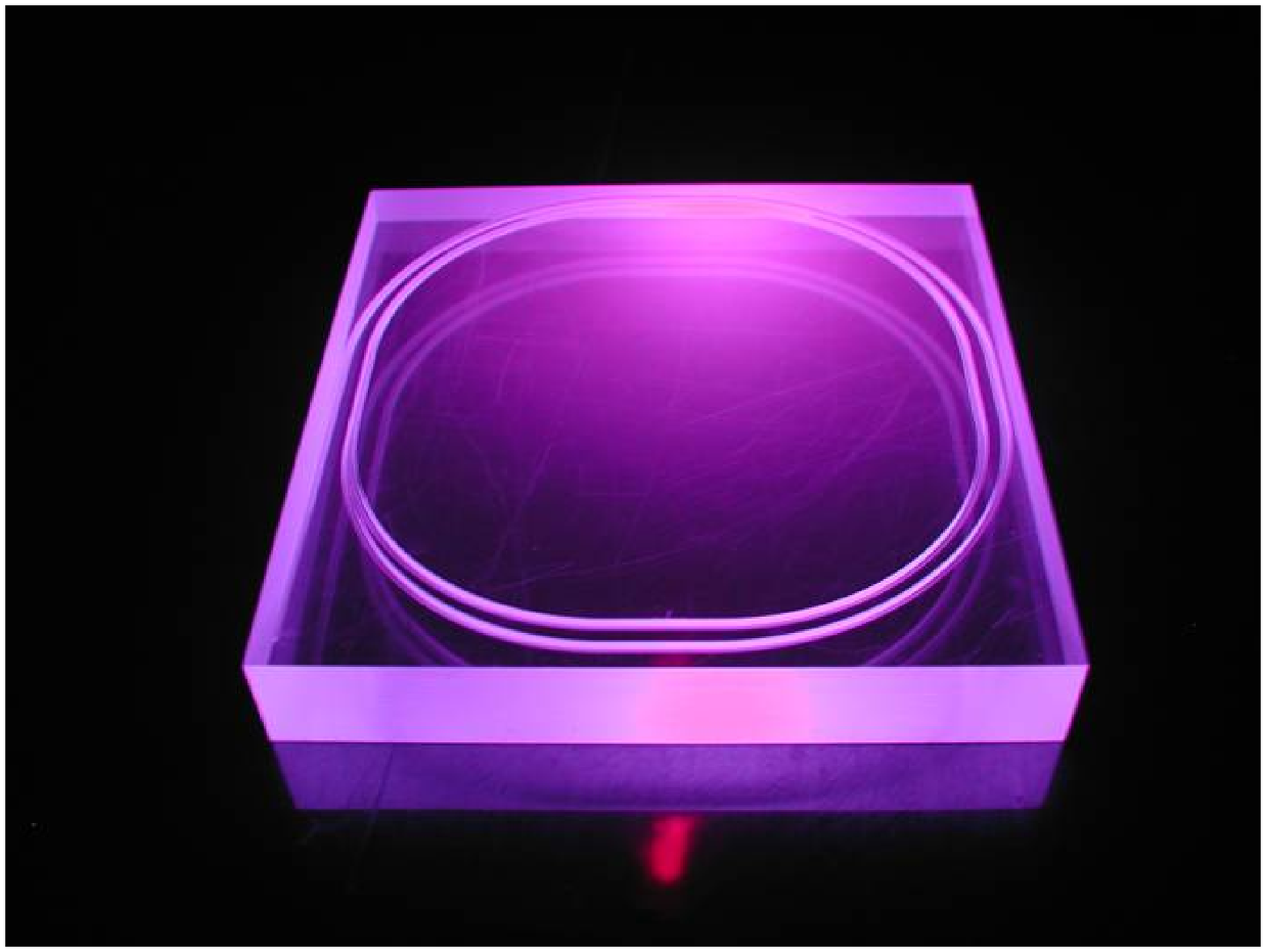}
\caption{CPR tile carved with a two-loop spiral groove path.}
\label{cpr_til}
\end{figure}

The new CPR detector is made of 20-mm thick scintillator tiles (Fig.~\ref{cpr_til})
read out through a 1-mm diameter WLS fiber embedded into a groove carved on the surface of each tile.
The groove cut inside the scintillator's surface has a 2-loop spiral design with a cross-sectional 
keyhole shape in order to maximize light collection.
After exiting the tile, each WLS fiber is spliced to a clear fiber, which terminates in a plastic connector at the module's edge.
Optical cables, approximately 5~m long, then transport the light to a 16-channel R5900 Hamamatsu {\it photomultiplier tube} (PMT) 
located in the back of the calorimeter wedge.
Each detector channel (i.e. one tile) is read out by one PMT pixel,
for a total of 2,592 (480) readout channels in the CPR (CCR) detector.

The CCR detector is replaced by a similar detector read out through the same technique. 
The CCR is located behind a 10 radiation length tungsten bar, which limits the scintillator thickness to 5~mm
and covers the uninstrumented regions, also called ``cracks'', 
in the azimuthal angle $\phi$ present between the calorimeter wedges. 
Ten tiles, approximately 5~cm wide, cover each $\phi$-crack with the same $\eta$-segmentation as that
of the central calorimeter of 10 towers per wedge. 
One WLS fiber is embedded into a straight groove in each tile.

\begin{figure}[h]
\centering
\includegraphics[width=3.0in]{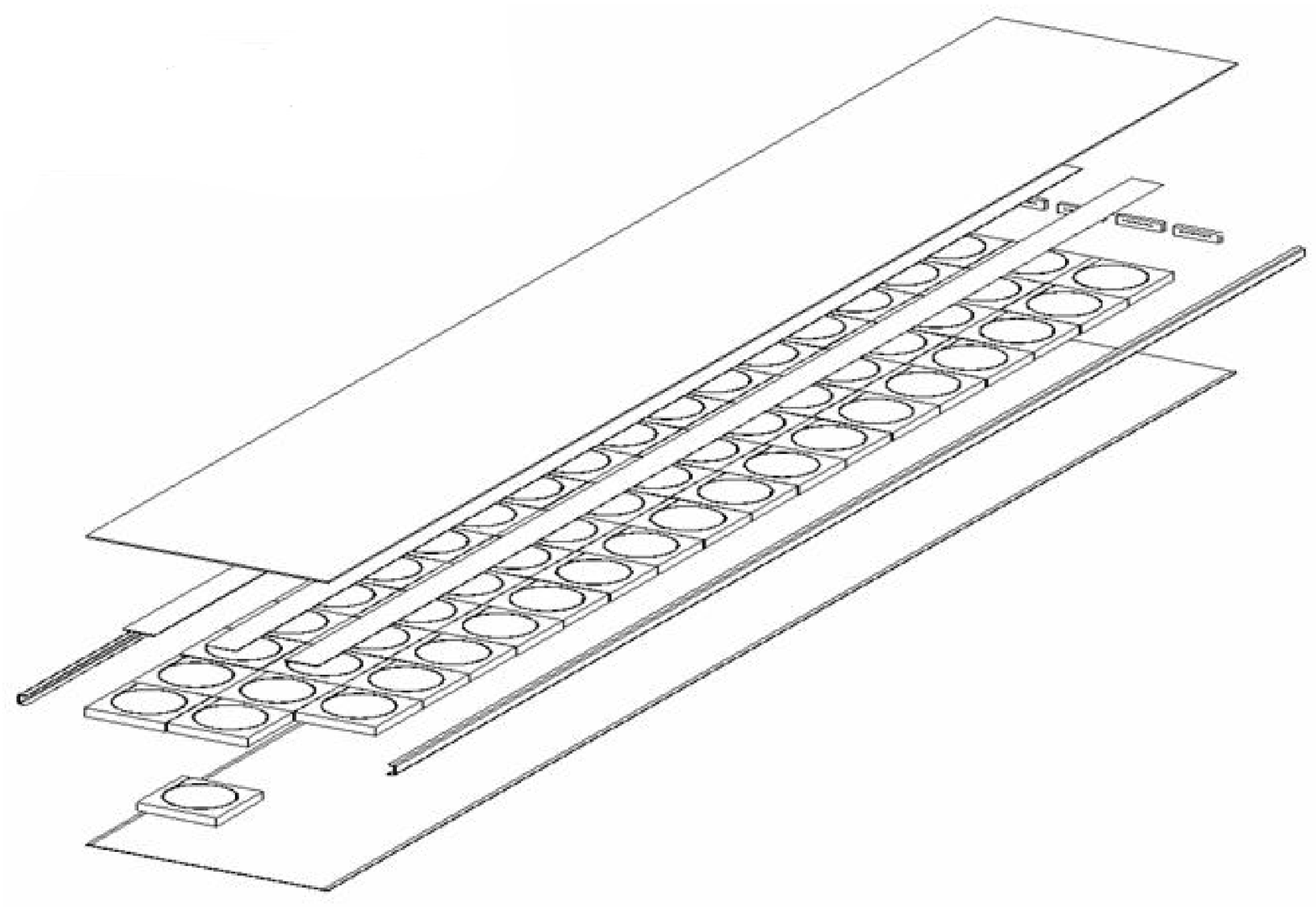}
\caption{Schematic drawing of one CPR upgrade module. 
Tiles are arranged in a 3x18 array, visible here before being sandwiched between the two cover plates.
Modules are oriented longitudinally to the beam.}
\label{cpr_mod}
\end{figure}

Both the CPR and the CCR detectors use the solenoid coil and the tracking material as a radiator.
The calorimeter wedges are arranged in two rings of 24 wedges each, that make contact at $z = 0$.
The entire detector consists of a total of 48 CPR (and 48 CCR) modules,
with each module covering the front face of one calorimeter wedge.
A set of 54 tiles forms one CPR module (Fig.~\ref{cpr_mod}) and is assembled in an aluminum shell, sealed to be light-tight.
The old CPR detector had a poorer segmentation with only 16$\times$2 towers for each calorimeter wedge~\cite{old}.
In the new CPR modules,
a continuous array of 3$\times$18 tiles (12.5$\times$12.5~cm$^2$ each) spans the face of one calorimeter wedge.
The front-end electronics uses the same readout modules previously employed by the old detector, with only minor modifications.

\section{Tests and detector performance}
\PARstart{I}{n} order to optimize the final design of the detector, several tests were performed to compare different kinds of
scintillators, fibers, and groove shapes.
As the new detector is placed in the space previously occupied by the old one, the dimensions are fixed and the
studies were mostly aimed at maximizing the light yield, in terms of the number of {\it photoelectrons} (pe) per MIP detected.

The new CPR detector was designed with the goal of being sensitive to the measurement of single particles,
which requires a minimum signal of 5 pe/MIP.
Unlike the old CPR, the new detector has no dead regions in $\eta$, thus allowing complete coverage.
On the other hand, the CCR is only designed to cover the cracks and detect early particle showers.
Preliminary tests indicated a light yield of 7~pe/MIP at the tile exit, 
which allows extension of the measurement of the electromagnetic shower energy to the crack regions.

Different tile/fiber configurations have been investigated and the results compared.
Cosmic rays were used to test the tile response to MIPs.
In order to increase the sampling of the light from the scintillator tile, a fiber is inserted into a groove whose 
path allows loading of multiple loops.
Two different groove shapes have been compared: keyhole and square grooves. 
In order to provide good optical coupling between fiber and scintillator tile,
optical cement was used in some tests.

\begin{figure}[h]
\centering
\includegraphics[width=3.0in]{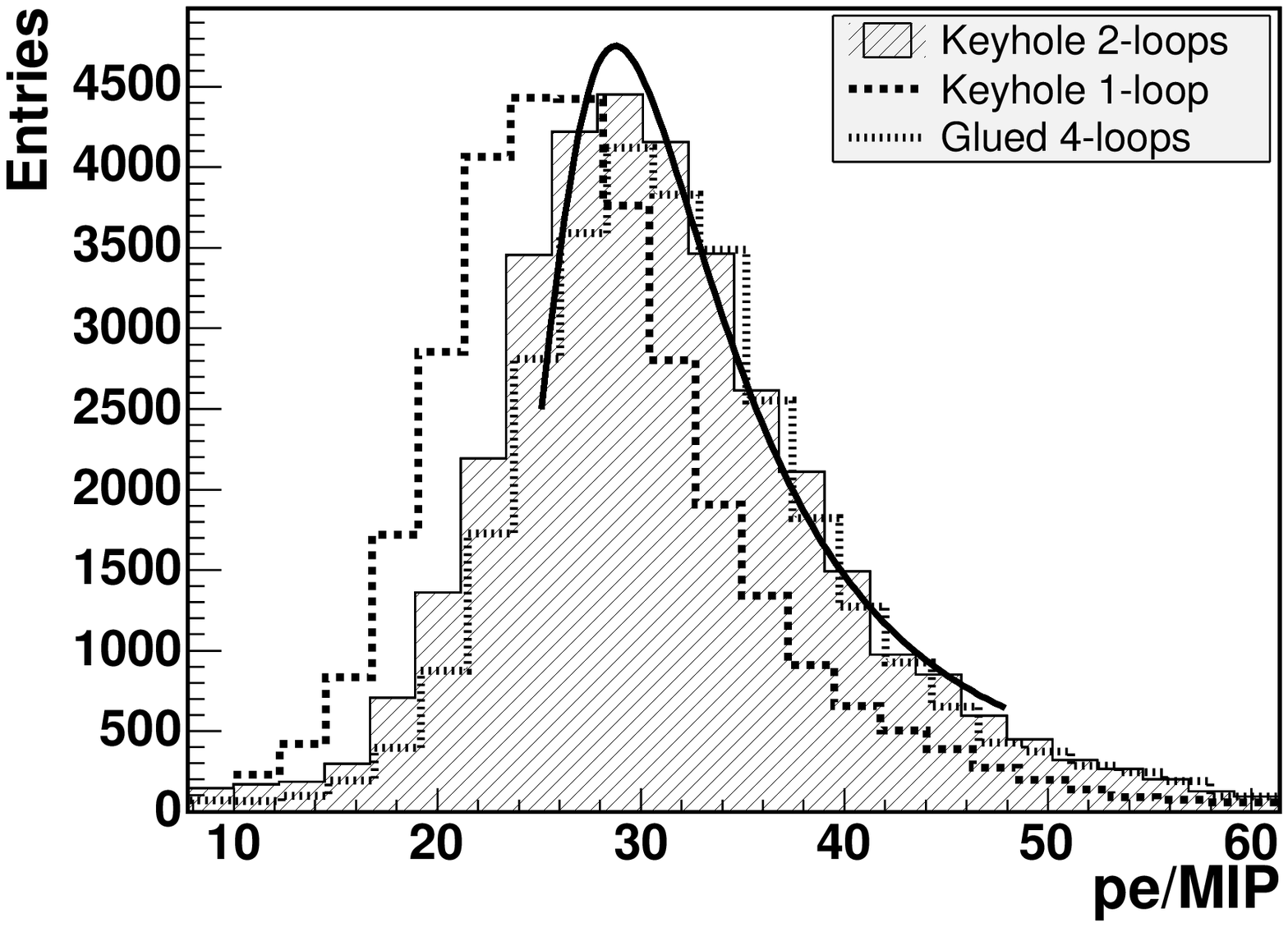}
\caption{Comparison of different methods of fiber/tile configurations. 
Light yield is optimized when the fiber is inserted in a 2-loop keyhole groove.}
\label{tile_comp}
\end{figure}

In Figure~\ref{tile_comp}, the light yield from a 2-loop spiral keyhole groove (solid shaded histogram and fit) 
compares well to the light yield from
a fiber glued in a 4-loop square groove (dotted histogram).
The keyhole-shaped groove allows capture of light from all sides while mantaining, at the same time,
the fiber inserted in place without the use of glue, thus
simplifying the assembly procedure.
In order to maximize the light yield, the scintillator tiles are also polished on all sides and wrapped in an aluminized mylar foil, 
acting as a reflector and separator between adjacent tiles. 

\begin{figure}[h]
\centering
\includegraphics[width=3.0in]{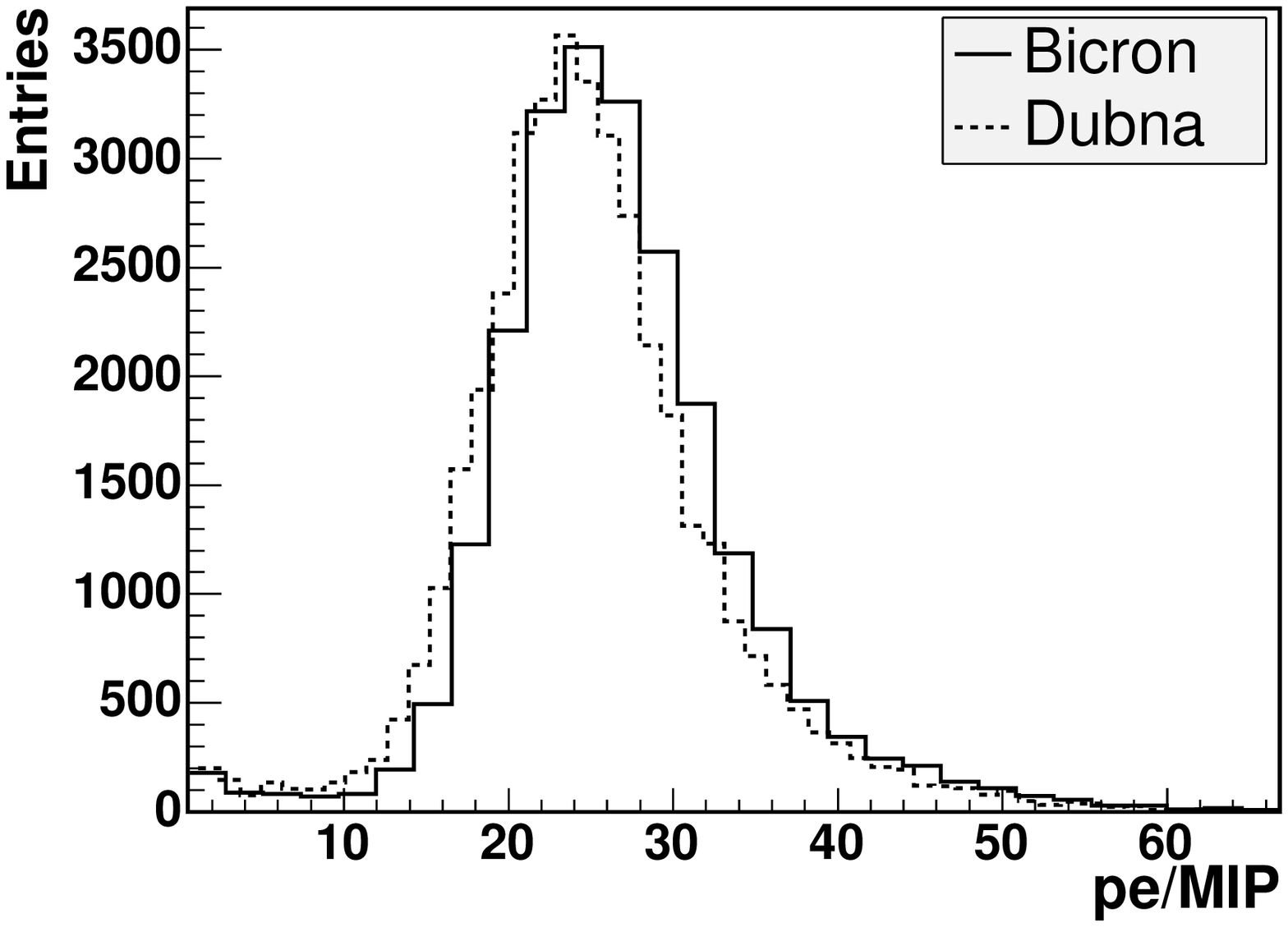}
\caption{Light yield for two different scintillator tiles. 
In this test, a non-mirrored WLS fiber by PolHiTech with a 4-loop configuration was used. 
Light yields from Bicron 408 and Dubna tiles are equal within 5\%.}
\label{cpr_dubna}
\end{figure}

Two different scintillator tiles were used for comparison, ``Dubna''~\cite{dubna} and Bicron 408~\cite{bicron}. 
Light yields obtained were similar within 5\% (Fig.~\ref{cpr_dubna}). 
Dubna tiles are prepared with polystyrene, a material with radiation hard properties, and were chosen in the final design.

Owing to the small size of the scintillator tile, the light travels a relatively short distance to the WLS fibers.
The light yield may be increased by using a scintillator with a high dopant concentration and a shorter attenuation length ($\sim$2-3~m).
Uniformity of light yield response within tiles was measured to be approximately 5\%.
WLS fibers manufactured by Kuraray~\cite{kuraray} and PolHiTech~\cite{polhitech} were compared, and
the attenuation length was measured to be approximately 5~m for both.
Kuraray fibers were chosen for their previous good performance in other CDF sub-detectors.
The light yield from multiple fiber loops inserted into the groove reaches a plateau
when the increase of light collection is compensated by the attenuation length of the fiber. 
The choice of two loops for the Kuraray WLS fibers is a compromise between assembly and light yield optimization.
The new CPR reads out only one end of the fiber, but much of the light transmitted in the other 
direction is recovered ($\sim$30\%) by mirroring that end of the fiber.
The Kuraray multi-clad Y11 WLS fiber embedded in a 2-loop spiral keyhole groove 
yields $\sim 30$~pe/MIP at the tile exit 
and is the preferred choice for the final detector.

\begin{figure}[h]
\centering
\includegraphics[width=3.3in]{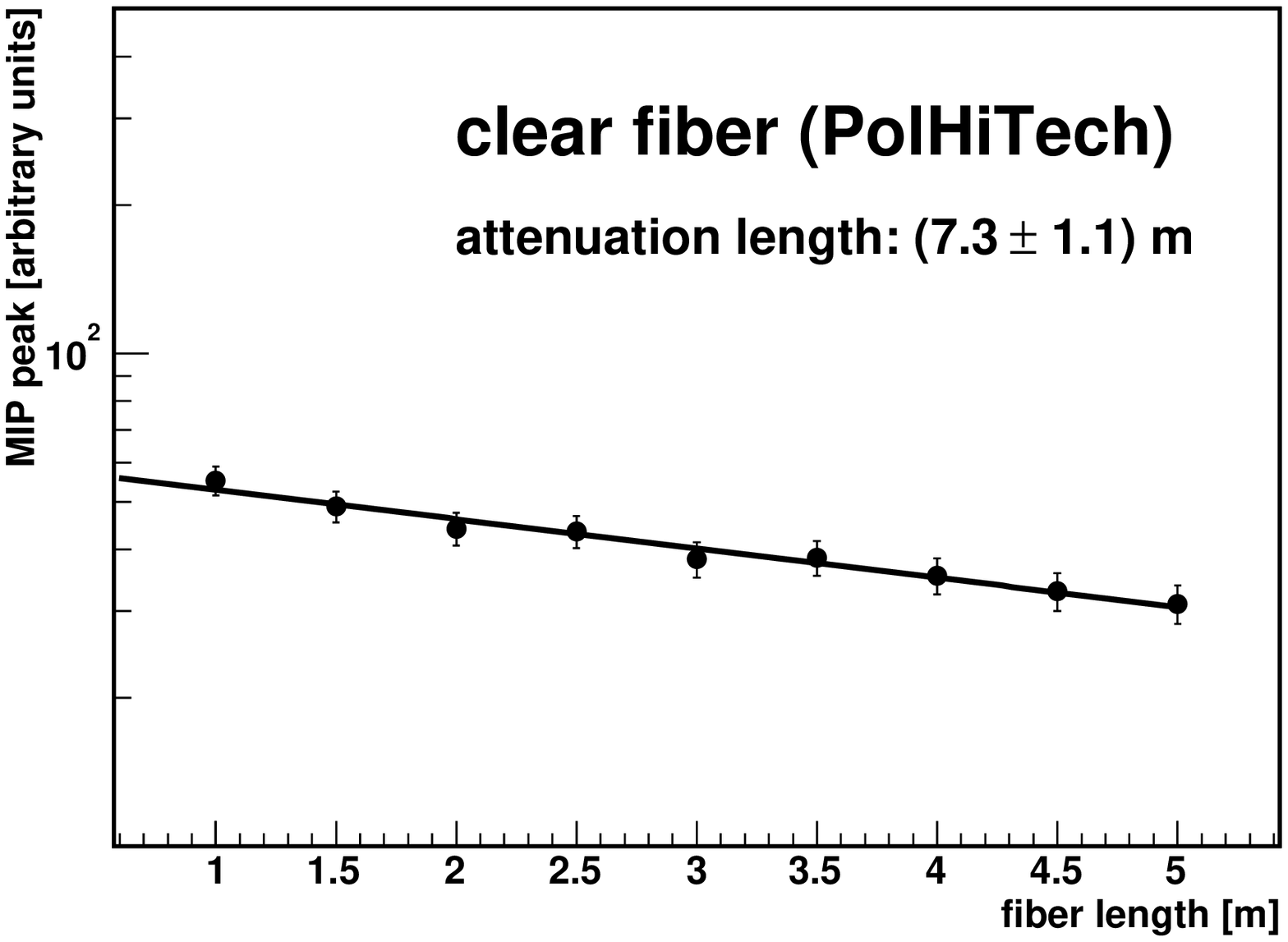}
\caption{Relative light yield response measured as a function of the length of one PolHiTech production clear fiber.}
\label{cpr_att}
\end{figure}

The clear fibers that transport the light to the multi-channel Hamamatsu PMTs are manufactured by PolHiTech
and their attenuation length was measured to be $\sim 7$~m, conforming to specifications (Fig.~\ref{cpr_att}).
More than 20~km were needed for the entire project. 

PMT uniformity has been tested prior to installation and conformed to specifications from Hamamatsu. 
The channel-to-channel variation has a maximum spread of 3:1; 
however, the fiber uses only the center of each pixel, and tests indicate a variation of only $\sim$10-15\% among pixels.
The crosstalk between direct neighbors is $\sim$1-2\%.
The high voltage supplied to the 192 PMTs is controlled by a CAEN~\cite{caen} crate (SY527) with eight 24-channel 932AN distributor boards.
Voltages have been adjusted to obtain approximately the same gain among PMTs.
Calibration constants have been included in a preliminary database to account for variation of the response among different channels
and will be further adjusted with colliding beam data, once the Tevatron resumes collisions.

\vspace*{0.6cm}

\section{Detector assembly}
\begin{figure}[h]
\centering
\includegraphics[width=3.4in]{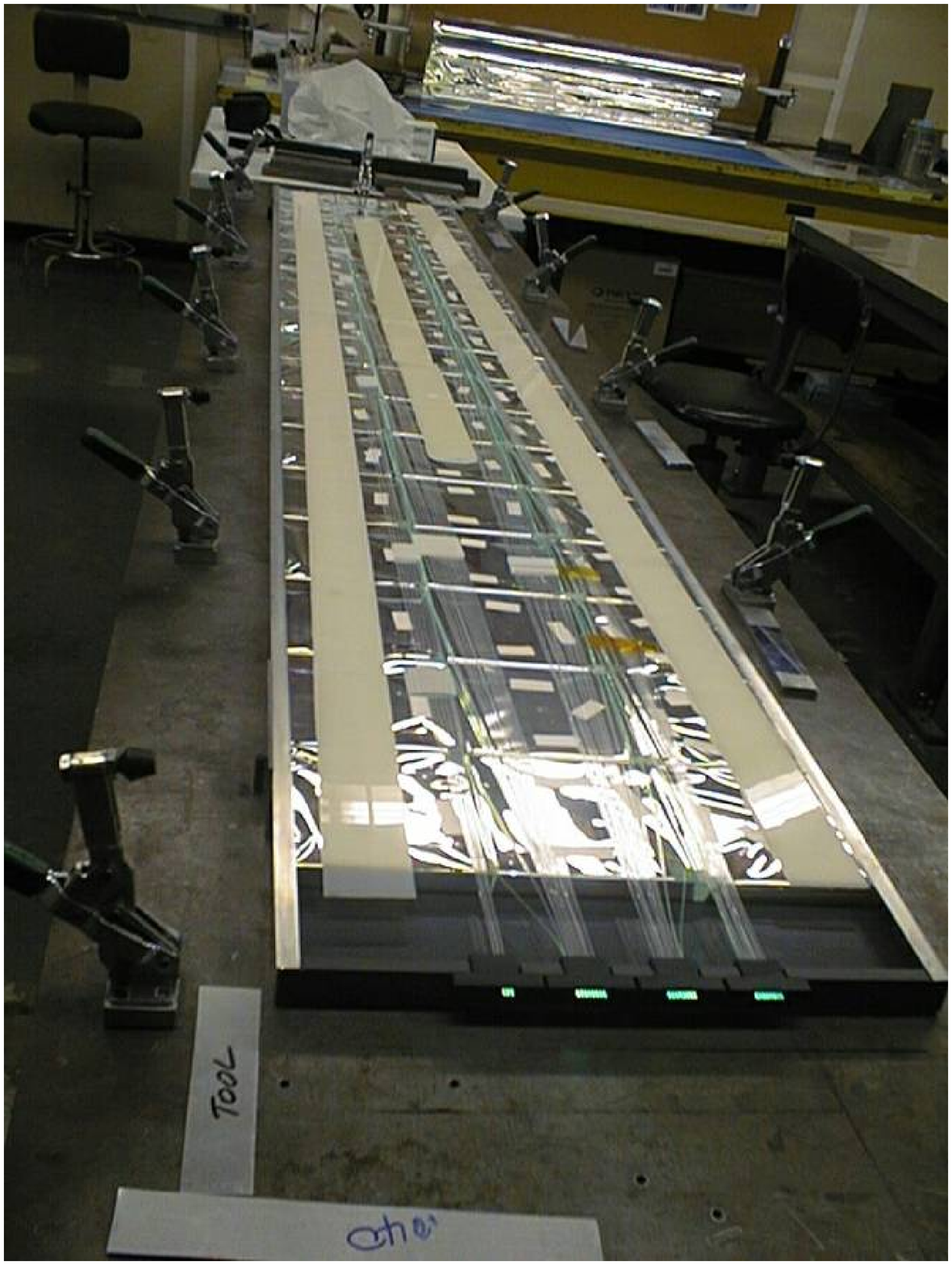}
\caption{Assembly of one CPR module: All 54 tiles are wrapped in aluminized mylar foil 
and the fibers routed to the four output connectors.}
\label{cpr_assembly}
\end{figure}

\begin{figure}[t]
\centering
\includegraphics[width=3.1in]{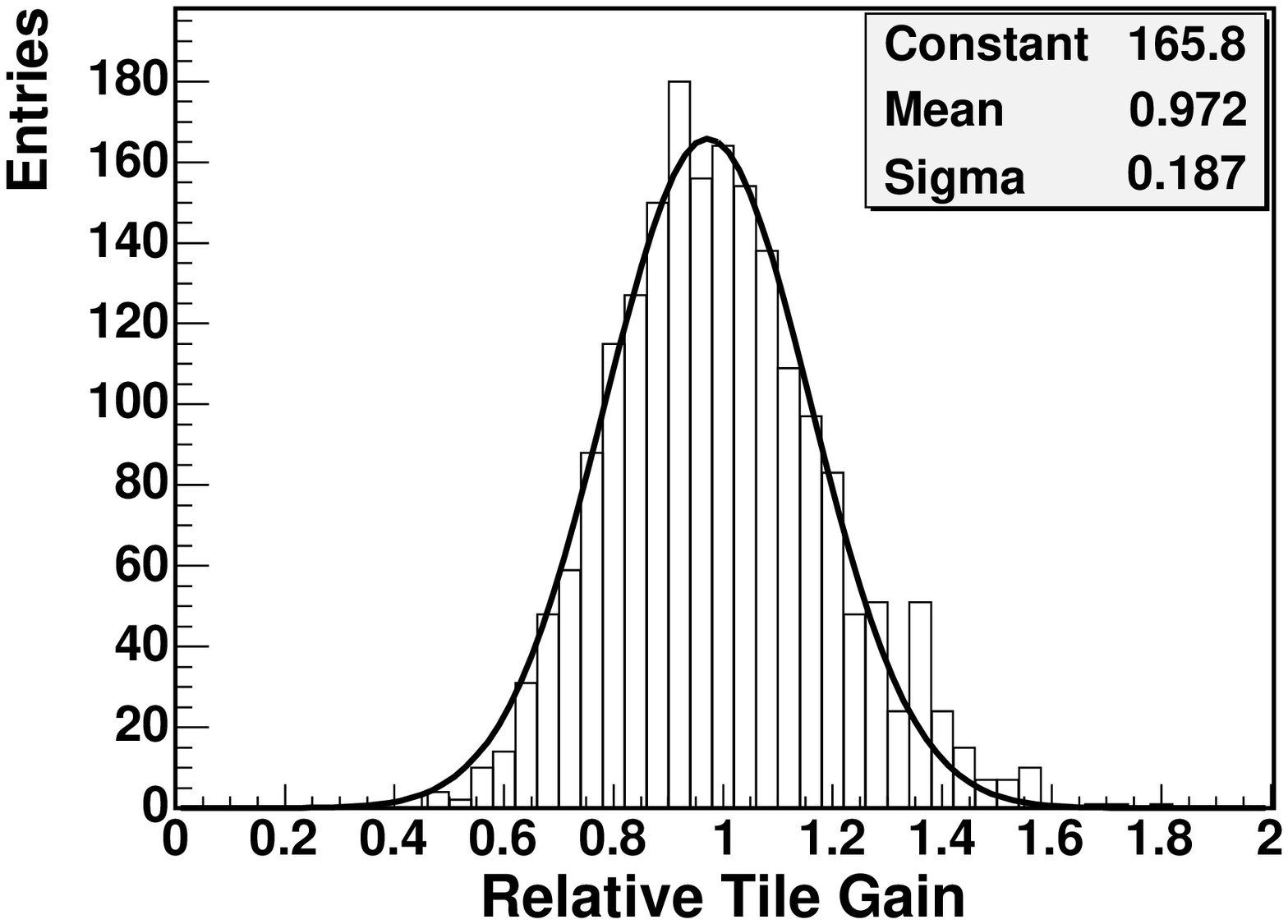}
\caption{Relative light yield response of the production scintillator tiles.}
\label{cpr_final}
\end{figure}

\PARstart{I}{n} preparation for detector assembly, 
the WLS fibers were spliced to clear fibers which were gathered into four groups per module and glued into plastic connectors.
The light transmission after splicing was measured to be $\sim$92-93\%, with good reproducibility and small uncertainties.
Prior to module assembly, both fibers and tiles were individually
tested for light yield uniformity using a radioactive source and visually inspected for damage.
During module assembly (Fig.~\ref{cpr_assembly}), each tile was first positioned in its location and 
the fiber then inserted in the keyhole groove and fixed in place, away from the tile's borders. 
The same procedure was repeated for all tiles, moving away from the connectors, until completion of the module. 
The fibers of the six farthest tiles were glued with optical cement to increase the light yield and compensate 
for a longer fiber path.
In order to ensure good quality control of module production, fibers were again individually checked for damage after assembly.
After closing the aluminum shell with light-tight sealant, 
each module was then scanned with a radioactive source and the light yield measured for all tiles. 
In Figure~\ref{cpr_final}, the relative light yield response for the tiles used in the production modules
has a spread of $\sim$19\%. 

\begin{figure}[t]
\centering
\includegraphics[width=3.2in]{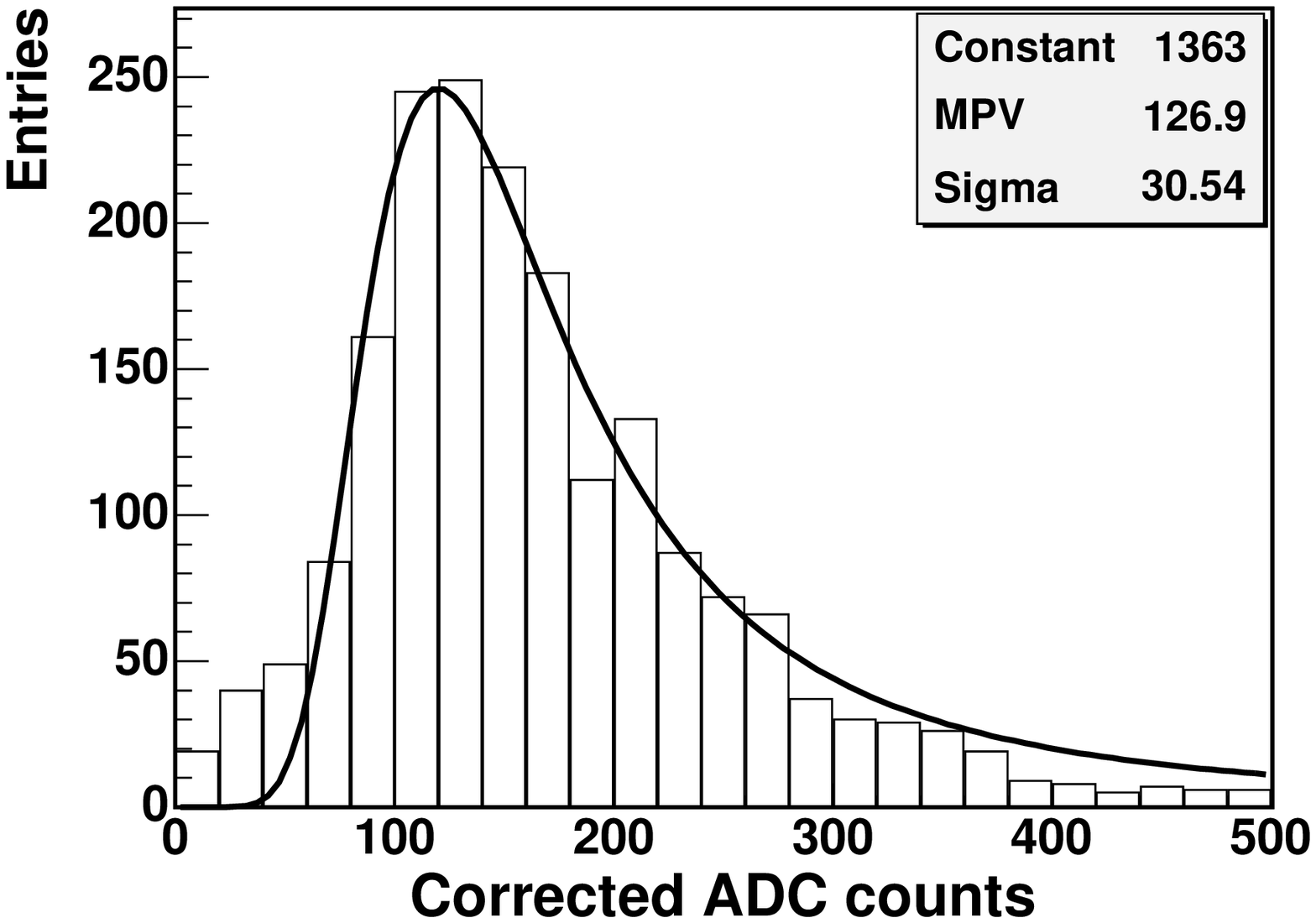}
\caption{ADC count distribution from cosmic ray data for the CPR modules after detector installation at CDF.
Ten ADC counts correspond to approximately 1~pe/MIP.
A light yield of 12 pe/MIP is estimated after the whole optical path.}
\label{mip_final}
\end{figure}

After installation, constant monitoring of detector response was performed using cosmic rays. 
The response of all installed modules was measured (Fig.~\ref{mip_final}) and
preliminary estimates indicate a light yield of 12 pe/MIP after the whole optical chain
(averaged over all the CPR modules), which is well above the design specifications of 5 pe/MIP.

\section{Conclusions}
\PARstart{T}{he} central preshower detector for the CDF experiment was designed and 
built after careful consideration of the physics issues that need to be addressed in the next few years at the Tevatron.
Particular attention was devoted to design studies aimed at optimizing the light yield and reliability. 
Detector production was completed smoothly and quality control was performed at various stages before, during, and after 
the construction phase. Detector installation took place during the fall 2004 in a timely fashion.
Tests indicate that the final detector meets and exceeds design requirements.
Indeed, preliminary measurements of light yield show a two-fold increase with respect to design goals. 
The calibration performed prior to installation allows valuable monitoring of detector performance.

\section*{Acknowledgments}
%
\PARstart{W}{e} thank the Fermilab staff and the technical staffs of the participating institutions for their vital contributions. This work was supported by the U.S. Department of Energy and National Science Foundation; the Italian Istituto Nazionale di Fisica Nucleare; the Ministry of Education, Culture, Sports, Science and Technology of Japan; the Natural Sciences and Engineering Research Council of Canada; the National Science Council of the Republic of China; the Swiss National Science Foundation; the A.P. Sloan Foundation; the Bundesministerium f\"ur Bildung und Forschung, Germany; the Korean Science and Engineering Foundation and the Korean Research Foundation; the Particle Physics and Astronomy Research Council and the Royal Society, UK; the Russian Foundation for Basic Research; the Comisi\'on Interministerial de Ciencia y Tecnolog\'{\i}a, Spain; in part by the European Community's Human Potential Programme under contract HPRN-CT-2002-00292; and the Academy of Finland.


\begin{thebibliography}{1}
\bibitem{timing} Another upgrade project for the CDF calorimeter includes the addition of timing information 
to both central and plug electromagnetic calorimeters, and
is needed to filter out cosmic ray and beam related backgrounds.
This project is described in more detail in Ref.~\cite{timing_ref}.
\bibitem{old} Each module of the old CPR gas detector had a structure of 16 strips, running longitudinally 
across five calorimeter towers and segmented in the azimuthal angle $\phi$. 
Each calorimeter wedge was covered by two modules, thus creating a dead region between the two.
\end{thebibliography}
\end{document}